\documentclass{jpp}
\usepackage{subeqn}
\usepackage{epsfig}

\let\realverbatim\verbatim
\let\realendverbatim\endverbatim
\renewenvironment{verbatim}
  {\par\addvspace{6pt plus 1pt minus 2pt}\realverbatim}
  {\realendverbatim\addvspace{6pt plus 1pt minus 2pt}}

\ifprodtf \else
  \checkfont{eurm10}
  \iffontfound
    \IfFileExists{upmath.sty}
      {\typeout{^^JFound AMS Euler Roman fonts on the system,
                   using the 'upmath' package.^^J}%
       \usepackage{upmath}}
      {\typeout{^^JFound AMS Euler Roman fonts on the system, but you
                   don't seem to have the}%
       \typeout{'upmath' package installed. JPP.cls can take advantage
                 of these fonts,^^Jif you use 'upmath' package.^^J}%
       \providecommand\umu{\umu}%
      }
  \else
    \providecommand\umu{\mu}%
  \fi
\fi

\ifprodtf \else
  \checkfont{msam10}
  \iffontfound
    \IfFileExists{amssymb.sty}
      {\typeout{^^JFound AMS Symbol fonts on the system, using the
                'amssymb' package.^^J}%
       \usepackage{amssymb}%

      }{}
  \else
  \fi
\fi

\ifprodtf \else
  \IfFileExists{amsbsy.sty}
    {\typeout{^^JFound the 'amsbsy' package on the system, using it.^^J}%
     \usepackage{amsbsy}}
    {}
\fi

\newcommand{\sig}{\sigma_{\rm ex}}
\newcommand{\be}{\begin{equation}}
\newcommand{\ee}{\end{equation}}

\newcommand{\eqa}{\begin{eqnarray}}
\newcommand{\eqe}{\end{eqnarray*}}
\newcommand{\eqnu}{\begin{eqnarray}}
\newcommand{\eqne}{\end{eqnarray}}

\newcommand{\eq}[1]{Eq. (\ref{#1})}

\newcommand{\eeq}{\end{eqnarray}}

\newdefinition{definition}[theorem]{Definition}

\title[Journal of Plasma Physics]
{An analytic model of plasma-neutral coupling in the heliosphere plasma}

\author[D. Shaikh]
{D\ls A\ls S\ls T\ls G\ls E\ls E\ls R \ns S\ls H\ls A\ls I\ls K\ls H\ls$^{1,2}$
\thanks{\tt Email:dastgeer.shaikh@uah.edu} \ls B. \ls D\ls A\ls S\ls G\ls U\ls P\ls T\ls A\ls$^2$
\thanks{\tt Email:brahmananda.dasgupta@uah.edu}}
\affiliation{$^1$Department of Physics and \\
$^2$Center for Space Physics and Aeronomic Research (CSPAR),\\
University of Alabama at Huntsville, Huntsville, AL 35805. USA.}
\date{May 29 2010}
\pubyear{2010} % only needed when year of publication is not current year.
\volume{00}
\part{0}
\pagerange{\pageref{firstpage}--\pageref{lastpage}}
\doi{S0963548301004989}

\begin{document}

\label{firstpage}
\maketitle

\begin{abstract}
We have developed an analytic model to describe coupling of plasma and
neutral fluids in the partially ionized heliosphere plasma
medium. The sources employed in our analytic model are based on a
$\kappa$-distribution as opposed to the Maxwellian distribution
function. Our model uses the $\kappa$-distribution to analytically
model the energetic neutral atoms that result in the heliosphere
partially ionized plasma from charge exchange with the protons and
subsequently produce a long tail which is otherwise not describable by
the Maxwellian distribution. We present our analytic formulation and
describe major differences in the sources emerging from these two
distinct distributions.

\end{abstract}

%\tableofcontents

\section{Introduction}
With the Voyager spacecraft now in the heliosheath (see Fig 1), the in
situ character of the solar wind plasma can be explored. Surprisingly,
the supersonic solar wind plasma, probed by the ACE/WIND/Cluster
spacecrafts near 1 AU (Astronomical Unit), depicts an entirely
different character when contrasted with the Voyager I and 2
observations in the heliosheath region (typically beyond 84 AU)
(Goldstein et al 1995, Burlaga et al 2005, 2006, 2008, 2009; Stone et
al 2005; Decker et al 2005; Richardson et al 2008, Zank 1999). Figure
1 shows an idealized cartoon reflecting our current understanding
based on theory, simulations and modeling together with
observations. Little is known about the physical processes that govern
the intricate multiscale (associated with waves, structures,
turbulence) interactions outlined in Fig 1. The supersonic solar wind
(SW) plasma interacts with local interstellar medium (LISM) neutral
hydrogen (H) gas through charge exchange leading to the creation of
energetic pick up ions (PUI).  The SW is decelerated, compressed and
heated at a shock, the termination shock (TS), across which it
develops small scale turbulence (Shukla 1978, Shaikh \& Zank 2008,
2010, Shaikh 2010, Shaikh et al 2006, Mendonca \& Shukla 2007).  In
the heliosheath region, the nonlinear structures, such as magnetic
hole and humps are found (Burlaga et al 2008, 2009). The SW protons
continue to interact with neutrals via charge exchange to produce
significant number of pick up ions. Both in the supersonic and
subsonic SW (at least in the outer heliosphere) the pressure
associated with the PUIs exceeds that of the solar wind protons
(Burlaga et al 2009).  Both Voyager 1 and 2 are reporting a number of
puzzling observations that were not anticipated by existing analytic
or simulation models. An intriguing example is that of magnetic field
distribution. The latter is lognormal in supersonic solar wind,
whereas it exhibits a Gaussian distribution in subsonic
heliosheath. Surprisingly, Voyager 2 indicates that the magnetic field
distribution is lognormal in the subsonic heliosheath plasma. The
source of this apparent discrepancy in the magnetic field
distributions reported by Voyagers 1 \& 2 in the heliosheath is not
known. Another example is that of plasma in heliosheath which is
compressed, turbulent and it is an admixture of waves, fluctuations
and magnetic structures (magnetic hole/hump, see section 2 for
details) (Burlaga 2006, 2009). The effect of PUIs on the formation and
evolution of nonlinear magnetic structures, waves and fluctuations in
outer heliosphere and the heliosheath plasma are an open
question. These issues continue to pose severe challenges to our
understanding of the heliosheath plasma.

\begin{figure}[t]
\includegraphics[width=13cm]{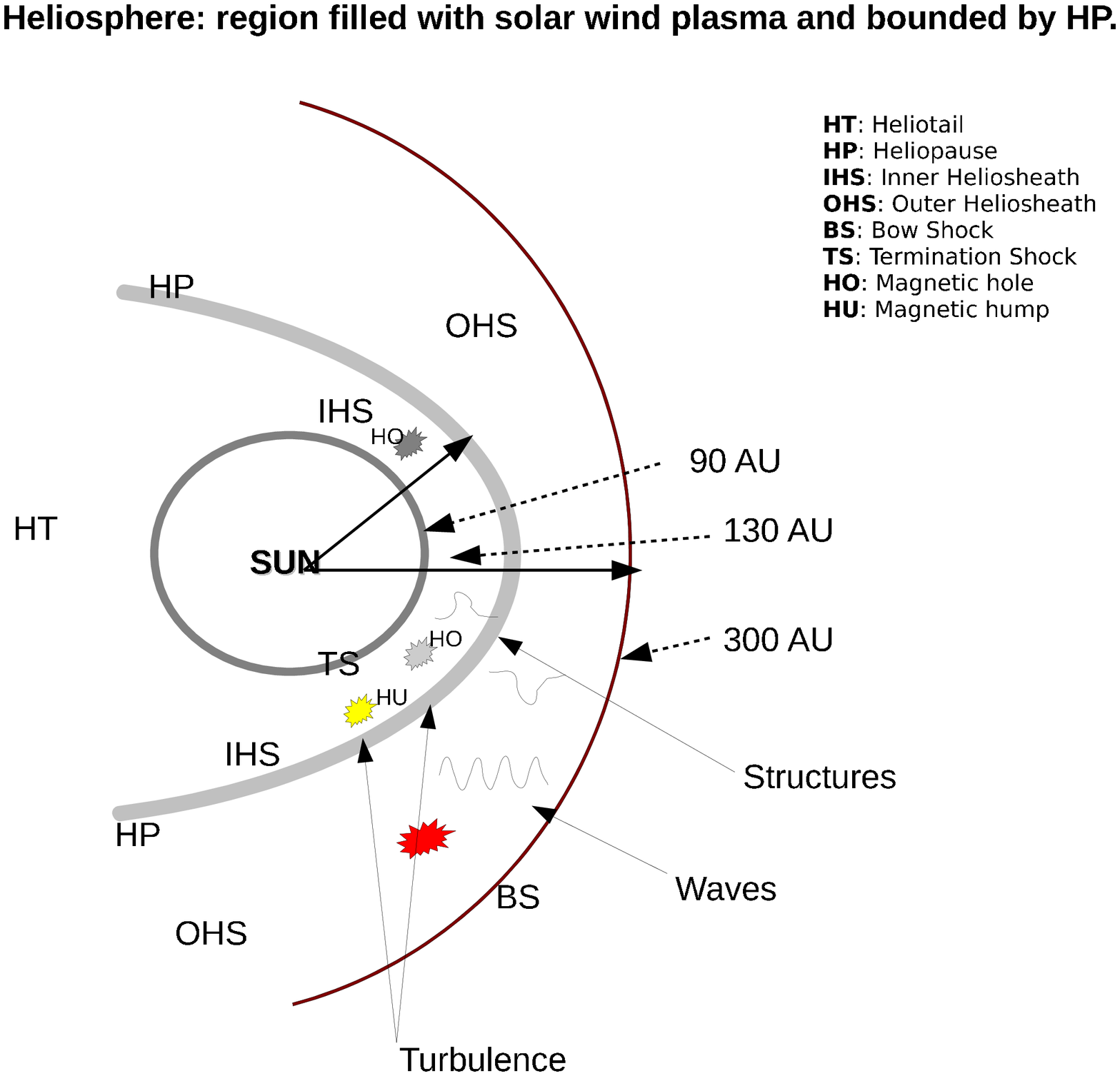}% Here is how to importEPS art
\caption{Schematic overview of different regions in the global heliosphere. The solar
wind emanating from the Sun propagates outward and interacts with
partially ionized interstellar gas predominantly via charge exchange,
and creates pick up ions (PUIs). At the termination shock (TS), the
supersonic SW decelerated, heated, compressed becoming subsonic, in
the heliosheath, and again interacts with interstellar neutrals via
charge exchange before it reaches heliopause (HP). The subsonic SW
flows down into the heliotail. Magnetic structures such as magnetic
holes/humps are observed in heliosheath plasma. During its journey
from the Sun to the HP, the solar wind plasma develops multitude of
length and time scales that interact with the partially ionized
interstellar gas, TS, and nonlinear structures develop in a complex
manner. }
\label{fig1}
\end{figure}

Although there exists wealth of in situ measurements by the Voyager
spacecrafts, they do not provide much information about the global
structure of the heliosphere interactions. For instance, the coupling
of plasma protons with the interstellar neutral atoms has
traditionally been done through Maxwellian sources. However, careful
studies have revealed that the distribution of hydrogen neutral (after
charge exchanging, they turn into energetic neutral atoms, ENA) does
not exactly follow a Maxwellian functions.  Recently, Prested et
al. (2008) used a $\kappa$-distribution for the ENA parent population
to obtain ENA maps. The advantage of using this distribution, as
opposed to a Maxwellian, is that it has a power-law tail, and is
therefore capable of producing ENA’s at suprathermal energies.  Now
there has been an increasing consensus that the plasma and neutral
fluids follow nearly kappa distribution (Heerikhuisen et al 2008).

A realistic modeling of the heliosheath plasma, one that includes a
self-consistent treatment of the PUIs, is therefore critically
important and essential to our understanding of the highly variable
heliosphere plasma.  The central them of this paper is therefore to
model complex coupling between plasma and neutral fluids via
$\kappa$-distribution as oppose to the Maxwellian distribution. Note
here that the $\kappa$ distribution modifies the charge exchange
interactions in fluid equations. The kappa-distribution emphasizes
charge exchange by high temperature protons. We will investigate the
effects of the $\kappa$-distribution in heliosphere plasma turbulence
for single fluid plasma-neutral coupled turbulence models.

In section 2, we describe $\kappa$-distribution for neutral and plasma
distribution and derive sources for the complex coupling interactions
between the two distinct fluids. Section 3 describes complete source
terms for the coupling interactions.  Finally, a summary is presented
in section 5.

\section{Plasma neutral coupling via $\kappa$-distribution source}

The charge exchange terms can be obtained from the Boltzmann transport
equation that describes the evolution of a neutral distribution
function $f({\bf r},{\bf v},t)$ in a six-dimensional phase space
defined respectively by position and velocity vectors ($x, v_x, v_y,
v_z$) at each time $t$. Here we follow Pauls et al. (1995) in
computing the charge exchange terms, based on $\kappa$-distribution
functions, from various moments of the Boltzmann equation. The
Boltzmann equation for the neutral distribution contains a source term
proportional to the proton distribution function $f_p$ and a loss
term proportional to the neutral distribution function $f_n$.
\begin{eqnarray}\label{prodp}
&&\frac{\partial }{\partial t}f_p({\bf r},{\bf v},t)+{\bf
v}\cdot\nabla f_p({\bf r},{\bf v},t)+ \frac{{\bf
F}}{m}\cdot\nabla_{v}f_p({\bf r},{\bf v},t)=
\nonumber\\
&&f_n({\bf r},{\bf v},t)\int f_p({\bf r},{\bf v_p},t)\left|{\bf
v_p}- {\bf v}\right|\sigma_{ex}(v_{rel})d^3{\bf v_p}-\nonumber\\
&&f_p({\bf r},{\bf v},t)\int f_n({\bf r},{\bf v_n},t)\left|{\bf
v_n}- {\bf v}\right|\sigma_{ex}(v_{rel})d^3{\bf v_n};
\end{eqnarray}
where, $\sigma_{ex}(v_{rel})$ is the charge exchange cross section.
The charge exchange parameter has a logarithmically weak dependence on
the relative speed ($v_{rel}=|{u}_p- {v}_n|$) of the neutrals and the
protons through $\sig = [(2.1-0.092 \ln (v_{rel})) 10^{-7} cm]^2$
(Fite et al 1962).  This cross-section is valid as long as energy does not
exceed $1eV$, which usually is the case in the inner/outer
heliosphere. Beyond $1eV$ energy, this cross-section yields a higher
neutral density. This issue is not applicable to our model and hence
we will not consider it here.  The density, momentum, and energy of
the thermally equilibrated Maxwellian proton and neutral fluids can be
computed from \eq{prodp} by using the zeroth, first and second
moments $\int f_{\xi} d^3\xi, \int m{\bf \xi} f_{\xi} d^3\xi$ and
$\int m\xi^2/2 f_{\xi} d^3\xi$ respectively, where $\xi={u}_p$ or
${v}_n$. Since charge exchange conserves the density of the proton and
neutral fluids, there are no sources in the corresponding continuity
equations. We, therefore, need not compute the zeroth moment of the
distribution function.  Computing directly the first moment
from \eq{prodp}, we obtain the neutral fluid momentum equation.

A similar evolution equation can be written for the
neutral distribution function $f_n({\bf r},{\bf v_n},t)$. We
consider the case where both $f_p({\bf r},{\bf v_p},t)$ and
$f_n({\bf r},{\bf v_n},t)$ are given by a $\kappa$ distribution of
the following type:
\begin{equation}\label{kappap}
f_p({\bf r},{\bf v_p}) = \frac{n_p}{\pi^{\frac{3}{2}}v_{T_p}^3}
\frac{\Gamma(\kappa+1)} {\kappa^{\frac{3}{2}}\Gamma\left(\kappa-
\frac{1}{2}\right)} \left[1+\frac{({\bf v_p}-{\bf U_p})^2}{\kappa
v_{T_p}^2}\right]^{-(\kappa +1)};
\end{equation}

\begin{equation}\label{kappan}
f_n({\bf r},{\bf v_n}) = \frac{n_n}{\pi^{\frac{3}{2}}v_{T_n}^3}
\frac{\Gamma(\kappa+1)} {\kappa^{\frac{3}{2}}\Gamma\left(\kappa-
\frac{1}{2}\right)} \left[1+\frac{({\bf v_n}-{\bf U_n})^2}{\kappa
v_{T_n}^2}\right]^{-(\kappa +1)}.
\end{equation}

First we evaluate the following integral:
\begin{equation}\label{betap1}
\beta_p({\bf r},{\bf v},t)= \sigma_{ex}(v_{rel})\int f_p({\bf
r},{\bf v_p},t)\left|{\bf v_p}- {\bf v}\right|d^3{\bf v_p};
\end{equation}
where $\sigma_{ex}(v_{rel})$ is taken out of the integral, as it
varies slowly with respect to $(v_{rel})$. The integral
(\ref{betap1}) is fully written as

\begin{equation}\label{betap2}
\beta_p({\bf r},{\bf v},t)= \sigma_{ex}(v_{rel})
\frac{n_p}{\pi^{\frac{3}{2}}v_{T_p}^3} A_\kappa \int
\left[1+\frac{({\bf v_p}-{\bf U_p})^2}{\kappa
v_{T_p}^2}\right]^{-(\kappa +1)}\left|{\bf v_p}- {\bf
v}\right|d^3{\bf v_p};
\end{equation} with $$ A_\kappa=\frac{\Gamma(\kappa+1)}
{\kappa^{\frac{3}{2}}\Gamma\left(\kappa- \frac{1}{2}\right)}.$$ We
write,
$${\bf v_p}-{\bf U_p} = ({\bf v_p}-{\bf v}) -({\bf U_p}-{\bf v})$$
and define new variables as
$${\bf V}=({\bf v_p}-{\bf v})/\sqrt{\kappa v_{T_p}^2};\quad
{\bf x}=({\bf U_p}-{\bf v})/\sqrt{\kappa v_{T_p}^2};$$
with the new variables, the integral in Eq (\ref{betap2}) becomes,

\begin{equation}\label{betap3}
\beta_p({\bf r},{\bf v},t)= \sigma_{ex}(v_{rel})
\frac{n_p}{\pi^{\frac{3}{2}}v_{T_p}^3} (\kappa v_{T_p}^2)^{3/2}
\sqrt{\kappa v_{T_p}^2}A_\kappa \int_{-\infty}^\infty
\left[1+({\bf V}-{\bf x})^2\right]^{-(\kappa +1)}V  d^3{\bf V};
\end{equation}
where we have used,
$$|{\bf v_p}-{\bf v}| =\sqrt{\kappa v_{T_p}^2} V; \quad d^3{\bf
v_p}=(\kappa v_{T_p}^2)^{3/2}d^3{\bf V}$$ and the constant before
the integral in Eq (\ref{betap3}) is,

$$\sigma_{ex}(v_{rel})\frac{n_p}{\pi^{\frac{3}{2}} v_{T_p}^3}
(\kappa v_{T_p}^2)^{3/2}\sqrt{\kappa
v_{T_p}^2}~\frac{\Gamma(\kappa+1)} {\kappa^{\frac{3}{2}}\Gamma
\left(\kappa- \frac{1}{2}\right)}=\sigma_{ex}
\frac{n_pv_{T_p}}{\pi^{\frac{3}{2}}}\frac{\sqrt{\kappa}~\Gamma(\kappa+1)}
{\Gamma \left(\kappa- \frac{1}{2}\right)}.$$
We now proceed to evaluate the integral Eq (\ref{betap3}). In spherical
coordinate,
\[d^3{\bf V}= V^2 dv \sin\theta ~d\theta ~d\phi, \]
 where
$\theta$ is the angle between {\bf V} and {\bf x}, after performing
the $\phi$ integration, with $\mu  =\cos\theta$

\begin{equation}\label{integ1}
I=2\pi\int^\infty_0 V^3 dV\int^1_{-1}d\mu(1+V^2-2Vx\mu+x^2)^{-
(\kappa+1)}
\end{equation}
This integration becomes,
$$I =2\left[\int^x_0z^2(1+z^2)^{-\kappa}dz+
x^2\int^x_0(1+z^2)^{-\kappa}dz +2x\int^\infty_x
z(1+z^2)^{-\kappa}dz\right],$$ with ${\bf x}=({\bf U_p}-{\bf
v})/\sqrt{\kappa v_{T_p}^2}$. We now proceed to determine the
explicit values of the above definite integrals. The first two
integrals are given in terms of the hypergeometric functions,
${_2F_1}$, which are

$$\int^x_0z^2(1+z^2)^{-\kappa}dz = \frac{x^3}{3}~{_2F_1}\left(
\frac{3}{2},\kappa;\frac{5}{2}; -x^2\right),$$

$$x^2\int^x_0(1+z^2)^{-\kappa}dz =x~{_2F_1}\left(
\frac{1}{2},\kappa;\frac{3}{2}; -x^2\right);$$ where the
Hypergeometric function ${_2F_1}(a, b; c; z)$ (with $a, b, c $ are
constant numbers and $z$ is the variable) is expressed as a power
series in $z$:

{{
\begin{eqnarray}\label{hypergeom1}
{_2F_1}(a, b; c; z)&=&1 + \frac{ab}{c}\frac{z}{1!}+
\frac{a(a+1)b(b+1)}{c(c+1)}\frac{z^2}{2!}\nonumber\\
&&\quad +\frac{a(a+1)(a+2)b(b+1)(b+2)}{c(c+1)(c+2)}\frac{z^3}{3!}+
...
\end{eqnarray}}}
Using Kummer identity for hypergeometric functions,
$${_2F_1}(a,b;c;z) = {_2F_1}(b,a;c;z)=
(1-z)^{-b}{_2F_1}[b,c-a;c;z/(z-1)] \eqno(\ref{hypergeom1}a)$$

$${_2F_1}\left(\frac{3}{2},\kappa;\frac{5}{2}; -x^2\right)=
(1+x^2)^{-\kappa}{_2F_1}\left(\frac{5-3}{2},\kappa;\frac{5}{2};
\frac{x^2}{1+x^2}\right)=(1+x^2)^{-\kappa}{_2F_1}\left(1,
\kappa;\frac{5}{2};\frac{x^2}{1+x^2}\right);
\eqno(\ref{hypergeom1}b)$$

Similarly,
$${_2F_1}\left(\frac{1}{2},\kappa;\frac{3}{2}; -x^2\right)=
(1+x^2)^{-\kappa}{_2F_1}\left( 1,\kappa;\frac{3}{2};
\frac{x^2}{1+x^2}\right)\eqno(\ref{hypergeom1}c)$$

The last integral can be evaluated easily

$$\int^\infty_x z(1+z^2)^{-\kappa}dz=\left.\frac{(1+z^2)^{-\kappa+1}}
{2(-\kappa+1)}\right|^\infty_x =\frac{(1+x^2)^{-\kappa+1}}
{2(\kappa-1)}\quad (\kappa ~{\rm must~be}>1)$$\\

Collecting the above terms, the integral $\beta_p({\bf r},{\bf
v})$ is,
\begin{eqnarray}\label{betap4}
\beta_p({\bf r},{\bf v},t)&=& \frac{2n_p\sigma_{ex}v_{T_p}}
{\sqrt{\pi\kappa}~x}\frac{\Gamma(\kappa+1)} {\Gamma \left(\kappa-
\frac{1}{2}\right)}(1+x^2)^{-\kappa}\left[x^2{_2F_1}\left(
1,\kappa;\frac{3}{2}; \frac{x^2}{1+x^2}\right)\right.\nonumber\\
&&+\left.\frac{x^2}{3}{_2F_1}\left(1,\kappa;\frac{5}{2};\frac{x^2}{1+x^2}\right)
+\frac{1+x^2}{\kappa-1}\right];\quad x= |{\bf U_p}-{\bf
v}|/\sqrt{\kappa v_{T_p}^2}\nonumber\\
\end{eqnarray}

An approximate value of the above expression in the two limits
$\sqrt{\kappa}~x \ll 1$ and $x\gg 1 $ can be obtained as follows:
\begin{equation}\label{betaps}
\sqrt{\kappa}~x \ll 1:\qquad\beta=\frac{2n_p\sigma_{ex}v_{T_p}}
{\sqrt{\pi\kappa}}\frac{\Gamma(\kappa+1)} {\Gamma \left(\kappa-
\frac{1}{2}\right)}\left[\frac{1}{\kappa-1}+\frac{({\bf U_p}-{\bf
v})^2}{3\kappa v_{T_p}^2}\right]
\end{equation}
\begin{equation}\label{betapl}
x \gg 1:\qquad\beta=n_p\sigma_{ex}\sqrt{\frac{4v_{T_p}
\Gamma^2(\kappa+1)} {\pi\kappa(\kappa-1)\Gamma^2 \left(\kappa-
\frac{1}{2}\right)}+({\bf U_p}-{\bf v})^2}
\end{equation}

Note that in an asymptotic limit, the Gamma functions for large argument
is
$$\lim_{\kappa\rightarrow\infty}\kappa^{b-a}\frac{\Gamma(\kappa+a)}
{\Gamma(\kappa+b)}\rightarrow 1,\quad \Rightarrow\quad
\lim_{\kappa\rightarrow\infty}\Gamma(\kappa+a)\simeq \kappa^a
\Gamma(\kappa)$$

\section{Complete expressions for the source terms}
To find the source terms, we take the moments of Eq (\ref{prodp}) by
multipling both sides with various powers of the velocity {\bf v}. The
zeroth order moment would contribute to the source term of the mass
continuity equation, the first order moment would contribute to the
source term of the momentum equation, the second order moment would
contribute to the source term of the energy equation. The moments for
the left hand terms of Eq (\ref{prodp}) are well known, so we shall
show the moments of the right hand terms, using kappa distribution for
$f_p$ and $f_n$. We derive full expression for the integrals on the
r.h.s of Eq (\ref{prodp}) by using the complete expression for the
integral for $ \beta_p({\bf r},{\bf v},t)$ given by Eq (\ref{betap4})
without any approximation.

Since the charge exchange process conserves the proton and neutral
density, there will be no source term for the mass continuity
term, so we need not calculate the zeroth moment. Hence we start
with the first moment of the right hand side of Eq (\ref{prodp}).
With

$$f_n({\bf r},{\bf v}) = \frac{n_n}{\pi^{\frac{3}{2}}v_{T_n}^3}
\frac{\Gamma(\kappa+1)} {\kappa^{\frac{3}{2}}\Gamma\left(\kappa-
\frac{1}{2}\right)} \left[1+\frac{({\bf v}-{\bf U_n})^2}{\kappa
v_{T_n}^2}\right]^{-(\kappa +1)}=f_n({\bf r},{\bf v}-{\bf U_n}),$$

$${\rm and}\hspace*{15ex} \beta_p({\bf r},{\bf v},t)\equiv \beta_p({\bf r},x,t)=
\beta_p({\bf r},{\bf U_p}-{\bf v})$$

The production term for \textbf{Momentum transport}, is,

\begin{eqnarray}\label{sourcemom}
Q_{MP}&=&\int_0^\infty d^3 v{\bf v}f_n({\bf r},{\bf v}-{\bf U_n})
\beta_p({\bf r},{\bf U_p}-{\bf v})\nonumber\\
&=&\int_0^\infty d^3 v{\bf U_n}f_n({\bf r},{\bf v}-{\bf U_n})
\beta_p[{\bf r},({\bf U_p}-{\bf U_n})-({\bf v}-{\bf U_n})]\nonumber\\
&&+\int_0^\infty d^3 v({\bf v}-{\bf U_n})f_n({\bf r},{\bf v}-{\bf
U_n}) \beta_p[{\bf r},({\bf U_p}-{\bf U_n})-({\bf v}-{\bf U_n})],\nonumber\\
&=&{\bf U_n}\int_0^\infty d^3 uf_n({\bf r},{\bf u}) \beta_p[{\bf
r},(\Delta{\bf U}-{\bf u})] +\int_0^\infty d^3 u{\bf u}f_n({\bf r},{\bf u})
\beta_p[{\bf r},(\Delta{\bf U}-{\bf u})];\nonumber\\
\end{eqnarray}where we used, ${\bf u} ={\bf v}-{\bf U_n};\quad \Delta{\bf
U}={\bf U_p}-{\bf U_n}$.

The production term for \textbf{Energy transport}, is,
\begin{eqnarray}\label{sourceenerg}
Q_{EP}&=&\int_0^\infty d^3 v|{\bf v}|^2f_n({\bf r},{\bf v}-{\bf
U_n})\beta_p({\bf r},{\bf U_p}-{\bf v})\nonumber\\
&=&\int_0^\infty d^3 v|{\bf v}-{\bf U_n}|^2f_n({\bf r},{\bf
v}-{\bf U_n})\beta_p({\bf r},{\bf U_p}-{\bf v})\nonumber\\
&&+2{\bf U_n}\cdot \int_0^\infty d^3 v{\bf v}f_n({\bf r},{\bf
v}-{\bf U_n})\beta_p({\bf r},{\bf U_p}-{\bf v})\nonumber\\
&&-U_n^2\int_0^\infty d^3 vf_n({\bf r},{\bf v}-{\bf
U_n})\beta_p({\bf r},{\bf U_p}-{\bf v})\nonumber\\
&=&\int_0^\infty d^3 v|{\bf v}-{\bf U_n}|^2f_n({\bf r},{\bf
v}-{\bf U_n})\beta_p({\bf r},{\bf U_p}-{\bf v})\nonumber\\
&&+2{\bf U_n}\cdot \int_0^\infty d^3 v({\bf v}-{\bf U_n})f_n({\bf r},
{\bf v}-{\bf U_n})\beta_p({\bf r},{\bf U_p}-{\bf v})\nonumber\\
&&+2 U_n^2\int_0^\infty d^3 vf_n({\bf r},{\bf v}-{\bf
U_n})\beta_p({\bf r},{\bf U_p}-{\bf v})\nonumber\\
 &&-U_n^2\int_0^\infty d^3 vf_n({\bf r},{\bf v}-{\bf
U_n})\beta_p({\bf r},{\bf U_p}-{\bf v})
\end{eqnarray}

As before, we introduce the variables ${\bf u} ={\bf v}-{\bf
U_n};\quad \Delta{\bf U}={\bf U_p}-{\bf U_n}$ and write
$\beta_p({\bf r},{\bf U_p}-{\bf v})=\beta_p({\bf r},({\bf U_p}
-{\bf U_n})-({\bf v}-{\bf U_n}))=\beta_p({\bf r},\Delta{\bf
U}-{\bf u})$. Expression (\ref{sourceenerg}) can be written as

\begin{eqnarray}\label{sourceenerg1}
Q_{EP}&=&\int_0^\infty d^3 uu^2f_n({\bf r},{\bf u})\beta_p({\bf
r},\Delta{\bf U}-{\bf u})+2{\bf U_n}\cdot \int_0^\infty d^3 u{\bf
u}f_n({\bf r},{\bf u})\beta_p({\bf r},\Delta{\bf U}-{\bf u})
\nonumber\\
&& +U_n^2\int_0^\infty d^3 uf_n({\bf r},{\bf u}) \beta_p({\bf
r},\Delta{\bf U}-{\bf u})\end{eqnarray}

\begin{figure}[t]
\includegraphics[width=13cm]{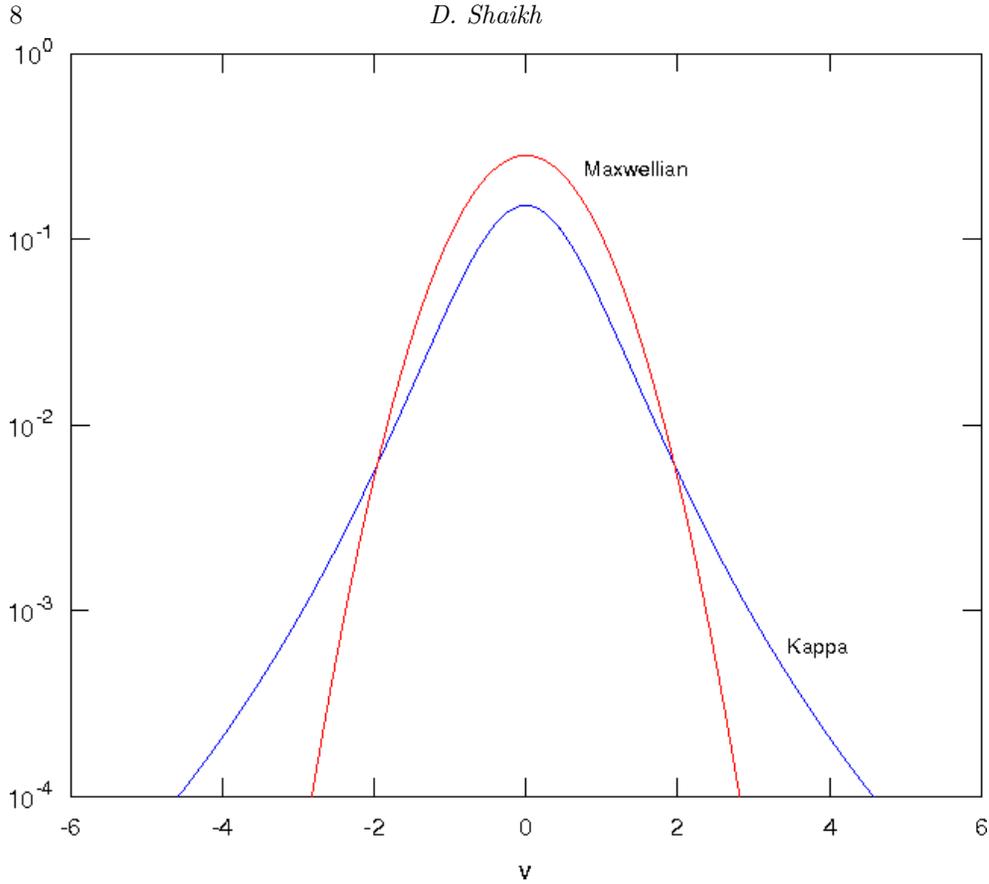}% Here is how to importEPS art
\caption{Comparision between the kappa and Maxwellian distribution
functions for the sources. Clearly, the tail of the distribution for
$\kappa$ function is long and wide as opposed to the Maxwellian
distribution. It is because of this feature, we have computed sources
based on a $\kappa$-distribution function. }
\label{fig2}
\end{figure}

\section{Summary and conclusion}
In summary, our major results are Eq (\ref{sourcemom}) \& Eq
(\ref{sourceenerg}). A tentative comparison of the sources based on
Maxwellian and kappa distribution functions is shown in Fig (2).  It
is evident from this figure that the sources based on a Maxwellian
distribution function falls off sharply and without any tail region.
This therefore excludes the energertic component of the neutral atoms
and hence is inappropriate for typical ENAs. By contrast, the sources
based on a kappa distribution function depicts a well-behaved tail
distribution that represents ENA distribution.

Our previous work in Shaikh \& Zank (2008) has shown that charge
exchange modes modify the helioshperic turbulence cascades
dramatically by enhancing nonlinear interaction time-scales on large
scales.  Thus the coupled plasma system evolves differently than the
uncoupled system where large-scale turbulent fluctuations are strongly
correlated with charge-exchange modes and they efficiently behave as
driven (by charge exchange) energy containing modes of helioshperic
turbulence.  By contrast, small scale turbulent fluctuations are
unaffected by charge exchange modes which evolve like the uncoupled
system as the latter becomes less important near the larger $k$ part
of the helioshperic turbulent spectrum.  The neutral fluid, under the
action of charge exchange, tends to enhance the cascade rates by
isotropizing the helioshperic plasma turbulence on a relatively long
time scale.  This tends to modify the characteristics of helioshperic
plasma turbulence which can be significantly different from the
Kolmogorov phenomenology of fully developed turbulence.  It remains to
be seen how these modified sources influence nonlinear turbulent
properties of the small scale helioshperic plasma fluctuations.

\section{Acknowledgment}
The partial support of NASA grants NNX09AB40G, NNX07AH18G, NNG05EC85C,
NNX09AG63G, NNX08AJ21G, NNX09AB24G, NNX09AG29G, and NNX09AG62G is
acknowledged.

%\begin{thereferences}{99}

\end{document}